\documentclass[prb,twocolumn,showpacs,preprintnumbers,amsmath,amssymb,superscriptaddress]{revtex4}

\usepackage{graphicx}
\usepackage{dcolumn}
\usepackage{bm}
\usepackage{mathrsfs}
\usepackage{subfigure}
\usepackage{natbib}

\begin{document}

\title{Conditions for non-monotonic vortex interaction in two-band superconductors}

\author{A. Chaves}
\altaffiliation{These authors contributed equally to this work.}
\affiliation{Departement Fysica, Universiteit Antwerpen,
Groenenborgerlaan 171, B-2020 Antwerpen, Belgium}
\affiliation{Departamento de F\'isica, Universidade Federal do
Cear\'a, 60455-900 Fortaleza, Cear\'a, Brazil}

\author{L. Komendov\'{a}}
\altaffiliation{These authors contributed equally to this work.}
\affiliation{Departement Fysica, Universiteit Antwerpen,
Groenenborgerlaan 171, B-2020 Antwerpen, Belgium}

\author{M. V. Milo\v{s}evi\'{c}}
\affiliation{Departement Fysica, Universiteit Antwerpen,
Groenenborgerlaan 171, B-2020 Antwerpen, Belgium}

\author{J. S. Andrade Jr.}
\affiliation{Departamento de F\'isica, Universidade Federal do
Cear\'a, 60455-900 Fortaleza, Cear\'a, Brazil}

\author{G. A. Farias}
\affiliation{Departamento de F\'isica, Universidade Federal do
Cear\'a, 60455-900 Fortaleza, Cear\'a, Brazil}

\author{F. M. Peeters}
\email{francois.peeters@ua.ac.be} \affiliation{Departement Fysica,
Universiteit Antwerpen, Groenenborgerlaan 171, B-2020 Antwerpen,
Belgium} \affiliation{Departamento de F\'isica, Universidade Federal
do Cear\'a, 60455-900 Fortaleza, Cear\'a, Brazil}

\date{\today}

\begin{abstract}
We describe a semi-analytic approach to the two-band Ginzburg-Landau
theory, which predicts the behavior of vortices in two-band
superconductors. We show that the character of the short-range
vortex-vortex interaction is determined by the sign of the normal
domain - superconductor interface energy, in analogy with the
conventional differentiation between type-I and type-II
superconductors. However, we also show that the long-range
interaction is determined by a {\it modified} Ginzburg-Landau
parameter $\kappa^*$, different from the standard $\kappa$ of a bulk
superconductor. This opens the possibility for non-monotonic
vortex-vortex interaction, which is {\it temperature-dependent}, and
can be {\it further tuned} by alterations of the material on the
microscopic scale.
\end{abstract}


\maketitle
\section{Introduction}
Multi-gap superconductivity arises when the gap amplitudes on
different sheets of the Fermi surface are radically disparate, e.g.
due to different dimensionality of the bands for the usual
phonon-mediated pairing, as is the case in MgB$_2$,\cite{MgB2} or
due to the repulsive pairing interaction, as it appears to be the
case in recently discovered iron-pnictides.\cite{pnic,pnic2} The
other examples of multi-gap materials include OsB$_2$, iron
silicides such as Lu$_2$Fe$_3$Si$_5$, chalcogenides (NbSe$_2$), but
also the conventional superconductors such as Pb when reduced to
nanoscale.\cite{nano}

In a strong magnetic field all superconducting condensates form
normal-metal voids, as an intermediate state before
superconductivity is fully destroyed. These normal domains tend to
{\it merge} in type-I superconductors in order to minimize their
positive surface energy, whereas in type-II superconductors they
have negative surface energy and {\it split} into quantized
vortices. However, in 2005 Babaev and Speight predicted the
so-called semi-Meissner state in two-band
superconductors,\cite{Babaev2005} the state with localized regions
of high and low vortex densities, arising from short-range repulsive
while long-range attractive vortex-vortex interaction. This vortex
behavior was recently visualized by Moshchalkov {\it et
al.},\cite{Mosh} in the form of stripes and clusters of vortices in
a single-crystal MgB$_2$. Such vortex configurations stemming from
the long-range attractive vortex behavior (see also Ref.
\onlinecite{Brandt} for review) are clearly very important in the
field of superconductivity, but they also present a bridge between
solid-state physics and soft condensed matter, where systems with
competing interactions are of abiding interest.\cite{reich}

To date the matter of competing vortex interactions in two-band
superconductors has not been conclusively settled although recent
years saw a surge of activities in this field. The original
prediction in Ref. \onlinecite{Babaev2005} concerns only the case when one
band is type-I and the other type-II, although it is unclear how
different types of behavior between bands in $k$-space (not real
space) can be discerned. Ref. \onlinecite{Babaev2010} demonstrated
such vortex behavior in systems where just one band is fully
superconducting, and the other superconducts only due to direct
coupling. Dao {\it et al.} found different types of possible
vortex-vortex interactions and several resulting exciting vortex
configurations, but did not provide a universal criterion to {\it a
priori} determine the type of vortex interaction.\cite{Dao} Finally
some authors expressed scepticism to nonmonotonic vortex
interaction; Geyer {\it et al.} showed that the normal metal/two-gap
superconductor surface energy close to $T_c$ depends just on a
single Ginzburg-Landau (GL) parameter $\kappa$, and thus only either
repulsive (type-II) or attractive (type-I) vortex-vortex interaction
is possible.\cite{Geyer} This point was later reenforced by Kogan
and Schmalian.\cite{kog}

\section{Methods and derivations}

\subsection{The Ginzburg-Landau formalism for two-band
superconductors}
In this paper we derive criteria for the appearance
of non-monotonic interaction of vortices in two-gap systems
described by the standard GL model. Our analysis is based on the
two-band GL theory, {\it but with correct microscopic parameters}
obtained either from theoretical band structure calculations or by
fitting the experimental penetration depth or specific heat data by
the so-called $\gamma$-model \cite{gammaModel}. We begin from the GL
energy functional, which comprises single-band contributions from
both condensates, the coupling term, and the energy of the magnetic
field in and around the sample:
\begin{eqnarray}
\label{GLfunctional} \mathcal{F}&=&\sum_{j=1,2}\alpha_j |\Psi_j|^2 +
\frac{1}{2} \beta_j |{\Psi_j}|^4 + \frac{1}{2m_j}\left|\left(
\frac{\hbar}{i}\nabla-\frac{2e}{c}\mathbf{A}\right)
\Psi_j\right|^2\nonumber\\&&-\Gamma(\Psi_1^*\Psi_2+\Psi_1\Psi_2^*)+\frac{(\mathbf
h-\mathbf H)^2}{8\pi}.
\end{eqnarray}
Here the two Cooper-pair condensates are described by the order
parameters $\Psi_1$ and $\Psi_2$, {\bf H} is the applied magnetic
field and {\bf h} the net one. The Josephson coupling term provides
the `minimal coupling', well described in literature. The
temperature enters the energy expression through $\alpha_{j=1,2}$,
linearly dependent on the temperature term $\tau=\ln
T_c/T\approx1-T/T_c$.\cite{zhit} The expansion leading to the GL
theory is strictly valid only in the immediate vicinity of $T_c$,
but we use this theory at somewhat lower temperatures as well,
arguing that GL theory qualitatively well describes important
physics away from $T_c$ (as was demonstrated at many prior
instances). Finally, it was shown in Ref. \onlinecite{kog}, that standard
two-band GL theory contains incomplete terms that estimate $\psi$
with precision to $\tau^{3/2}$. The authors reduce the theory by
eliminating latter terms, which results in a single coherence length
for both order parameters of a two-band superconductor. This is
however not a correct physical picture at low temperatures, and two
coherence lengths for the two-band superconductors can be recovered
even in the GL domain in the extended model of Ref.~\onlinecite{shan}.
Unfortunately, the latter model is presented in the absence of
magnetic field. To be able to capture all the essential physics, at
least qualitatively, we base our study on the compromise standard
GL model. Note however that our further explained semi-analytic
approach can be applied to {\it any improved form} of the energy
functional for two-band superconductors.

We next calculate the vortex-vortex interaction in a similar fashion
to Ref. \onlinecite{Babaev2005} but within a correct microscopic
framework. The parameters in Eq.~(\ref{GLfunctional}) can then be
expressed as: $\alpha_j = -N(0)n_j\chi_j = -N(0)n_j(\tau -
S_j\big/n_j\eta)$, $\beta_j = N(0)n_j/W^2$,
$m_j=3W^2\big/N(0)n_jv_j^2$ and $\Gamma=N(0)\lambda_{12}\big/\eta$,
where $\Lambda= \left|
\begin{array}{ccc}
\lambda_{11} & \lambda_{12} \\
\lambda_{21}=\lambda_{12} & \lambda_{22}
\end{array} \right|$ is the coupling
matrix with determinant $\eta$; $n_j$ ($N(0)$) denotes partial
(total) density of states, $v_j$ are the Fermi velocities in the two
bands, and $W^2=8\pi^2T_c^2\big/7\zeta(3)$. For details on constants
$S_j$ we refer to Ref. \onlinecite{kog}. This allows us to
technically define the coherence lengths $\xi_j=\frac{\hbar
v_j}{\sqrt{6}W}$ and penetration depths
$\lambda_j=\sqrt{\frac{3c^2}{16\pi N(0) e^2 n_j v_j^2}}$, as well as
the GL parameters $\kappa_j=\lambda_j/\xi_j$ of the two condensates,
as if they were independent. These are however just parameters of
the model, and are related only indirectly with the resulting
penetration depth and the healing lengths of the two order
parameters in the two-band material. Notice also that $\alpha_1$ and
$\alpha_2$ change sign at different temperatures. In particular,
close to $T_c$ both $\alpha_j$ are positive but the coupled system
is still superconducting. Such situation is already different from
the one studied in Ref. \onlinecite{Babaev2010}, where at least one
$\alpha_j$ was negative. The Ginzburg-Landau equations minimize the
functional from Eq.~(\ref{GLfunctional}), and read (in dimensionless
form)
\begin{subequations}
\begin{equation}
(-i\nabla- \mathbf A)^2 \Psi_1
-(\chi_1-|\Psi_1|^2)\Psi_1-\gamma\Psi_2 =0,
\end{equation}
\begin{equation}
(-i\nabla- \mathbf A)^2 \Psi_2
-\alpha(\chi_2-|\Psi_2|^2)\Psi_2-\frac{\gamma\kappa_2^2}{\kappa_1^2\alpha}
\Psi_1 =0,
\end{equation}
\begin{equation}
-\triangle \mathbf{A}=\kappa_1^{-2}j_1 + \alpha \kappa_2^{-2} j_2,
\end{equation}
\label{gleq}
\end{subequations}
where $j_{j}=\Re \left[ \Psi_j^* (-i\nabla -\mathbf A )\Psi_j
\right]$, $\alpha=(v_1/v_2)^2$, $\gamma = \Gamma\big/n_1N(0)$, both
order parameters are scaled to $W$, distances to $\xi_1$, and vector
potential to $hc\big/4e\pi\xi_1$.

\subsection{Long-range vortex interaction} In what follows, we
demonstrate the method to determine the asymptotic long-range
interaction of vortices, before going into fine details at short
vortex-vortex distances. In cylindrical coordinates, considering the
ansatz for one circular symmetric vortex $\Psi_j =
e^{i\theta}f_j(r)$, and substituting the gauge $\vec{A} =
a(r)\hat{\theta}/r$, we rewrite Ginzburg-Landau Eqs.~(\ref{gleq}a-c)
as
\begin{subequations}\label{eqCIRC}
\begin{equation}
\frac{d^2f_1}{dr^2} + \frac{1}{r}\frac{df_1}{dr} -
\frac{(a-1)^2}{r^2}f_1 + (\chi_1 - f_1^2)f_1 + \gamma f_2 = 0,
\end{equation}
\begin{equation}
\frac{d^2f_2}{dr^2} + \frac{1}{r}\frac{df_2}{dr} -
\frac{(a-1)^2}{r^2}f_2 + \alpha(\chi_2 - f_2^2)f_2 +
\frac{\gamma}{\alpha}\frac{\kappa_2^2}{\kappa_1^2}f_1 = 0,
\end{equation}
and
\begin{equation}
\frac{d^2a}{dr^2} - \frac{1}{r}\frac{da}{dr} -
(a-1)\left(\frac{f_1^2}{\kappa_1^2} +
\alpha\frac{f_2^2}{\kappa_2^2}\right) = 0.
\end{equation}\label{eqAFix}
\end{subequations}
For $r\rightarrow\infty$, $a$ converges to 1 and $f_j$ to a
constant $a_j$. The limit $r\rightarrow\infty$ leads to
the set of non-linear coupled equations for $a_j$:
\begin{subequations}\label{eqCIRClim}
\begin{equation}
(\chi_1 - a_1^2)a_1 + \gamma a_2 = 0,
\end{equation}
\begin{equation}
\alpha(\chi_2 - a_2^2)a_2 +
\frac{\gamma}{\alpha}\frac{\kappa_2^2}{\kappa_1^2}a_1 = 0.
\end{equation}
\end{subequations}
These can be decoupled by defining the ratio $\rho = a_1/a_2$,
which then obeys the fourth order equation
\begin{equation}
\frac{\gamma}{\alpha^2}\frac{\kappa_2^2}{\kappa_1^2}\rho^4 +
\chi_2\rho^3 - \chi_1\rho -\gamma = 0.
\end{equation}
Such an equation has a laborious analytical solution known as
Ferrari's method, which will not be presented here, but can be found
in Ref.~\onlinecite{Abramowitz}. From Eq. (\ref{eqCIRClim}), one obtains
the dependence of the constants $a_j$ on the ratio $\rho$ as
\begin{subequations}\label{a1a2}
\begin{equation}
a_1 = \sqrt{\frac{\gamma}{\rho} + \chi_1},
\end{equation}
\begin{equation}
a_2 =
\sqrt{\frac{\gamma}{\alpha^2}\frac{\kappa_2^2}{\kappa_1^2}\rho +
\chi_2}.
\end{equation}
\end{subequations}

In order to eliminate high order terms for large distances, we
must use auxiliary functions that approach zero as
$r\rightarrow\infty$, namely, $Q(r) = a(r) - 1$ and $ \sigma_j(r)
= f_j(r) - a_j$. Keeping only first order terms in these
functions, Eqs. (\ref{eqCIRC}) become
\begin{subequations}
\begin{equation}
\frac{d^2\sigma_1}{dr^2} + \frac{1}{r}\frac{d\sigma_1}{dr}
+\left(\chi_1 - 3a_1^2\right)\sigma_1 + \gamma \sigma_2 = 0,
\end{equation}
\begin{equation}
\frac{d^2\sigma_2}{dr^2} + \frac{1}{r}\frac{d\sigma_2}{dr} +
\alpha\left(\chi_2 - 3a_2^2\right)\sigma_2 +
\frac{\gamma}{\alpha}\frac{\kappa_2^2}{\kappa_1^2} \sigma_1 = 0,
\end{equation}
and
\begin{equation}
\frac{d^2}{dr^2}\left(\frac{Q}{r}\right) +
\frac{1}{r}\frac{d}{dr}\left(\frac{Q}{r}\right) -
\left(\frac{\xi_{1}^2}{\lambda^2}-\frac{1}{r^2}\right)\left(\frac{Q}{r}\right)
= 0,
\end{equation}\label{Asympt}
\end{subequations}
where we defined $\lambda^{-2} = (a_1/\lambda_1)^{2} +
(a_2/\lambda_2)^{2}$. The solution of Eq. (\ref{Asympt}c) is the
Modified Bessel function $Q(r) = \delta_3rK_1(r\xi_1/\lambda)$.
Similarly, if $\gamma = 0$, Eqs. (\ref{Asympt}a) and (\ref{Asympt}b)
are decoupled and easily identified as Modified Bessel equations,
whose solutions are $\sigma_1(r) = \eta_1K_0(\sqrt{2\chi_1}r)$ and
$\sigma_2(r) = \eta_2K_0(\sqrt{2\alpha\chi_2}r)$. On the other hand,
if $\gamma \neq 0$, the equations for $\sigma_j$ are still coupled
and, in order to decouple them, one must define the operator
$\hat{L}_2 = \nabla^2 + \alpha\left(\chi_2-3a_2^2\right)$, so that
$\hat{L}_2\sigma_2 = -(\gamma\kappa_2^2\big/\alpha\kappa_1^2)
\sigma_1$, and apply it on Eq. (\ref{Asympt}a), obtaining
\begin{equation}\label{quarticdiff} \nabla^2\nabla^2\sigma_1 + C_1
\nabla^2\sigma_1 + C_2 \sigma_1 = 0.
\end{equation}
Here $C_1 =
\left(\chi_1-3a_1^2\right)+\alpha\left(\chi_2-3a_2^2\right)$ and
$C_2 =
\alpha\left(\chi_2-3a_2^2\right)\left(\chi_1-3a_1^2\right)-\gamma^2\kappa_2^2\big/\alpha\kappa_1^2$.
The operator $\nabla^2$ for axially symmetric solutions has
eigenfunctions given by Bessel functions $J_0(\beta r)$ and
$Y_0(\beta r)$, with eigenvalue $-\beta^2$, or modified Bessel
functions $I_0(\beta r)$ and $K_0(\beta r)$, with eigenvalue
$\beta^2$. From these four eigenfunctions, only the latter satisfies
the condition that $\sigma_j$ must decay monotonically with $r$.
Substituting $\nabla^2 K_0(\beta r) = \beta^2K_0(\beta r)$ in Eq.
(\ref{quarticdiff}), one obtains
\begin{equation}
\beta^4 + C_1 \beta^2 + C_2 = 0,
\end{equation}
and
\begin{subequations}\label{sigmas}
\begin{equation}
\sigma_1 (r) =
\delta_1\cos(\omega)K_0(\beta_{-}r)-\delta_2\sin(\omega)K_0(\beta_{+}r),
\end{equation}
\begin{equation}
\sigma_2 (r) =
\delta_1\sin(\omega)K_0(\beta_{-}r)+\delta_2\cos(\omega)K_0(\beta_{+}r),
\end{equation}
where
\begin{equation}
\beta_{\pm} = \sqrt{\frac{-C_1 \pm \sqrt{C_1^2-4C_2}}{2}}.
\label{beta}
\end{equation}
\end{subequations}
Notice that in Eqs. (\ref{sigmas}), each $\sigma_j$ must contain the
Bessel functions for both $\beta_{\pm}$, in a combination that is
conveniently written in the form of a mixing angle
$\omega$.\cite{Babaev2010}
 In the $\gamma \rightarrow 0$ limit, one has
$\beta_- \rightarrow \sqrt{2\chi_1}$ and $\beta_+ \rightarrow
\sqrt{2\alpha\chi_2}$. Moreover, substituting Eqs. (\ref{sigmas})
in the differential equation (\ref{Asympt}a), one obtains
\begin{subequations}
\begin{equation}
\tan(\omega) = \frac{\gamma}{\beta_+^2 + (\chi_1-3a_1^2)},
\end{equation}
\end{subequations}
so that $\gamma \rightarrow 0$ leads to $\omega \rightarrow 0$
and, consequently, to $\sigma_1(r) \rightarrow
\eta_1K_0(\sqrt{2\chi_1}r)$ and $\sigma_2(r) \rightarrow
\eta_2K_0(\sqrt{2\alpha\chi_2}r)$, as expected.

The parameters $\delta_k$ ($\eta_k$) in the expressions for $Q(r)$,
$\sigma_1(r)$ and $\sigma_2(r)$ are unknown real constants that can
only be determined by fitting numerical solutions for Eqs.
(\ref{eqCIRC}) in analogy to what is done in Ref. \onlinecite{Kramer1971}.

Having the asymptotic form of the order parameters and the vector
potential, we now follow the standard procedure \cite{Speight1997}
for finding the vortex-vortex interaction in the $r\rightarrow
\infty$ limit, obtaining
\begin{equation}
E_{2B}(\textsf{r}) = \delta_3^2
K_0\left(\frac{\textsf{r}}{\lambda}\right)-\delta_1^2K_0
\left(\frac{\beta_-\textsf{r}}{\xi_{1}}\right)-
\delta_2^2K_0\left(\frac{\beta_+\textsf{r}}{\xi_{1}}\right),
\label{Eint2B}
\end{equation}
where the units are now explicitly shown. Here, we list the consequences of the above asymptotics. i)
Comparing Eq. (\ref{Eint2B}) to the one-band case
\cite{Mohamed2002}, where
\begin{equation}
E_{1B}(\textsf{r}) = \delta_4^2
K_0\left(\textsf{r}/\lambda_{1B}\right)-
\delta_5^2K_0\left(\sqrt{2}\textsf{r}/\xi_{1B}\right),
\label{Eint1B}
\end{equation}
shows that the lengthscale $\lambda^{-2} = (a_1/\lambda_{1})^{2} +
(a_2/\lambda_{2})^{2}$ is playing the role of an effective
penetration depth for the two-band superconductor in accordance with
Eq.~(60) in Ref.~\onlinecite{gur}, contrary to $\lambda^{-2} =
(1/\lambda_{1})^{2} + (1/\lambda_{2})^{2}$ used in Refs.
\onlinecite{Babaev2005} and \onlinecite{Wang2009} which holds only
in the (unrealistic) absence of coupling. ii) The parameters
$\delta_k$ are in general different from each other, but can be
calculated exactly in the Bogomol'nyi point for the two band system
as $\delta_1^2 = \delta_2^2 = 2\delta_3^2$. For $\gamma = 0$, the
choice of $\xi_{1} = \xi_{2} = 1$ and $\kappa_1 = \kappa_2 = 1$ in
the two-band case is thus analogous to the Bogomol'nyi point
$\kappa_{1B} = 1/\sqrt{2}$ for the single band case and,
accordingly, the long-range interaction must vanish (and change sign
for $\kappa_1 = \kappa_2 < 1$). This directly illustrates that
coupling of two (nominally) type-II condensates may lead to a type-I
behavior of the coupled system! iii) In Eq. (\ref{Eint1B}) for
single band superconductors, it is clear that if $\kappa_{1B} =
\lambda_{1B}/\xi_{1B} > 1/\sqrt{2}$ ($< 1/\sqrt{2}$), the
interaction potential $E_{1B} (\textsf{r})$ will be repulsive
(attractive). For two-band superconductors, Eq. (\ref{Eint2B}) shows
that the relevant parameters are $\kappa^*_{\pm} =
\frac{\beta_{\pm}\lambda}{\sqrt{2}~\xi_{1}}$, rather than the
nominal GL parameters $\kappa_j$ for each condensate. If either
$\kappa^*_{+}$ or $\kappa^*_{-}$ are below $1/\sqrt{2}$, the
long-range vortex interaction is attractive (type-I like). Eqs.
(\ref{eqCIRClim}) and (\ref{beta}) provide simple means to evaluate
this condition. iv) In the presence of coupling, the long-range
behavior of both $\sigma_j$ depends exponentially on \emph{the
smallest} of $\beta_-$ and $\beta_+$. Therefore, in the coupled
case, we can define not only a single penetration depth for both
bands, but also the order parameters for both condensates exhibit
{\it the same decay} at large distances which implies a joint
coherence length $\xi^*=\xi_{1}/\min(\beta_{+},\beta_{-})$.

\subsection{Surface energy and the short-range vortex interaction}
The analysis in the previous subsection brings us to the discussion
of the real criterion for the attractive/repulsive nature of the
vortex interaction. In the single-band case, the changing sign of
the normal domain - superconductor surface energy $E_S$ at the
Bogomol'nyi point is a correct criterion. However, in the two-band
case and for large vortex-vortex distance, the Bogomol'nyi point is
determined by a single valued
$\kappa^*=\min(\kappa^*_{+},\kappa^*_{-})=1/\sqrt{2}$, which is not
necessarily where the surface energy of the normal domain (vortex)
changes sign!

The sign of the energy of the interfaces between normal-metal
domains and the superconductor determines whether merging of
those domains is energetically favorable or not (i.e. if the
superconductor is type-I or type-II). In the case of vortices, the
smallest possible normal domains, the positive vortex-superconductor
surface energy therefore means that the vortices should repel (at
least at short distances) in order to avoid the formation of a giant
vortex. We here show how to calculate the normal-superconducting
interface energy and by that predict the type of the short-range
vortex-vortex interaction.

We follow a similar approach to that of Ref.~\onlinecite{Wang2009}, but
we take into account the Josephson coupling and the temperature
dependence of the Ginzburg-Landau (GL) parameters, within a correct
microscopical framework. Namely, we consider the interface between
normal and superconducting region as the $yz$-plane at $x = 0$ and
calculate the surface energy $E_S$ using the one-dimensional GL
functional at the thermodynamic critical field $H_{cc}$, which reads
\begin{align}\label{EqSurf}
E_S = \int_{-\infty}^{\infty}dx \Big\{2(\Psi_1'^2 + A^2\Psi_1^2) + (2\chi_1 -\Psi_1^2)\Psi_1^2 \nonumber \\
+\alpha\frac{\kappa_1^2}{\kappa_2^2}\left[2(\Psi_2'^2 + A^2\Psi_2^2) + \alpha(2\chi_2 -\Psi_2^2)\Psi_2^2\right] \nonumber \\
-2\gamma \Psi_1\Psi_2 + \left(H_{cc}-\sqrt{2}\kappa_1 A'\right)^2\Big\}  ,
\end{align}
where the gauge potential
is chosen as $\vec{A} = (0,A(x),0)$ and $\Psi_{j=1,2}$ are taken
real. The thermodynamic critical field of the coupled system $H_{cc}$ is obtained from the condition that the GL functional in
Eq.~(\ref{GLfunctional}) converges to zero for $H = H_{cc}$, leading
to
\begin{equation}
H_{cc}^2 = H_{c(1)}^2a_1^2(2\chi_1-a_1^2) +
H_{c(2)}^2a_2^2(2\chi_2-a_2^2) + 4\gamma H_{c(1)}^2a_1a_2.
\end{equation}
We then find $\Psi_i$ and $A$ that minimize $E_S$ by numerically solving the set of Euler-Lagrange equations for the functional in Eq.~(\ref{EqSurf}),
which are exactly the one-dimensional versions of Eqs. (\ref{gleq}a-c):
\begin{subequations}
\begin{equation}
\Psi_1'' = \frac{A^2}{2}\Psi_1 - \left(\chi_1 -
\Psi_1^2\right)\Psi_1-\gamma\Psi_2,
\end{equation}
\begin{equation}
\Psi_2'' = \frac{A^2}{2}\Psi_2 - \alpha\left(\chi_2 -
\Psi_2^2\right)\Psi_2
-\frac{\gamma\kappa_2^2}{\alpha\kappa_1^2}\Psi_1,
\end{equation}
\begin{equation}
A'' = \left(\frac{\Psi_1^2 }{\kappa_1^{2}} +
\alpha\frac{\Psi_2^2}{\kappa_2^{2}}\right)A.
\end{equation}
\end{subequations}
The boundary conditions in the normal state ($x \to -\infty$) and deep in
the superconducting state ($x \to \infty$) are $\psi_j
(x\to -\infty) = 0$, $A'(x \to -\infty)=1$,
$\psi_j'(x\rightarrow\infty) = 0$ and $A'(x\rightarrow\infty)=0$.

\subsection{Constrained GL equations for fixed vortices}

We supplement our argumentation by numerically obtained
vortex-vortex interaction potentials (in a similar fashion as in
Ref. \onlinecite{Chaves}). Since the problem of two vortices does
not have circular symmetry, we now consider the fixed-vortex ansatz
in Cartesian coordinates $\Psi_j =
e^{in_1\theta_1}e^{in_2\theta_2}f_j(x,y)$, describing two fixed
vortices with winding numbers $n_1$ and $n_2$, where $e^{i n_k
\theta_k}$ is written in Cartesian coordinates as
\begin{equation}
e^{i n_k \theta_k} = \left(\frac{x_k + iy_k}{x_k -
iy_k}\right)^{n_k/2},
\end{equation}
and $\vec{r}_{k} = (x_k, y_k, 0) $ is the in-plane position vector
with origin at the center of the vortex $k$. For the case of two
vortices separated by a distance $d$, we take $\vec{r}_{1} = (x -
d/2, y, 0) $ and $\vec{r}_{2} = (x + d/2, y, 0)$. With this ansatz,
the Euler-Lagrange equations for the energy functional in Eq.~(1) read (see also Ref. \onlinecite{Chaves})
\begin{subequations}\label{ELfix}
\begin{eqnarray}
\nabla^2f_1 - \left[\overline{X}^2 + \overline{Y}^2 +
2(A_x\overline{Y}-A_y\overline{X})+ \vec{A}^2\right]f_1 \nonumber \\
+ (\chi_1-f_1^2)f_1 + \gamma f_2 = 0, \label{EL31}
\end{eqnarray}
\begin{eqnarray}
\nabla^2f_2 - \left[\overline{X}^2 + \overline{Y}^2 +
2(A_x\overline{Y}-A_y\overline{X})+ \vec{A}^2\right]f_2  \nonumber \\
+ \alpha\left(\chi_2-f_2^2\right)f_2 + \frac{\gamma
\kappa_2^2}{\alpha \kappa_1^2} f_1 = 0, \label{EL32}
\end{eqnarray}
and
\begin{equation}
\vec{\nabla} \times \vec{\nabla} \times \vec{A} = - \left[ \vec{A}
- \frac{ n_1\hat{\theta}_1}{r_1} - \frac{n_2\hat{\theta}_2}{r_2}
\right]\left(\frac{f_1^2}{\kappa_1^2} +
\alpha\frac{f_2^2}{\kappa_2^2}\right) = 0, \label{EL4}
\end{equation}
\end{subequations}
where
\begin{equation}
\overline{X} = \frac{n_1 x_1}{r_1^2}+ \frac{n_2 x_2}{r_2^2},
\quad\quad\quad \overline{Y} = \frac{n_1 y_1}{r_1^2}+ \frac{n_2
y_2}{r_2^2}, \nonumber \label{eqx}
\end{equation}
and the angular unit vectors around each vortex are written as
$\widehat{\theta_k} = (-y_k/r_k, x_k/r_k, 0)$.

Eqs.~(\ref{ELfix}a-c) are thus the GL equations for the two fixed
vortices, and we solve them numerically by a relaxation method. The
obtained order parameter and vector potential are then substituted
back in the energy functional, yielding the energy $E(d)$ for the vortex
pair at distance $d$. Repeating this procedure for different
vortex-vortex separation, we obtain the interaction potential $\Delta E = E(d)- E(0)$ between vortices in the two-gap superconductor, as shown in
Fig.~\ref{fig1}(b-d).

\section{Results and discussion}
We now apply the techniques described in the previous section to
calculate (i) the asymptotic long-range GL parameter $\kappa^*$,
(ii) the normal domain - superconductor surface energy $E_S$, and
(iii) the full vortex-vortex potential (using the constrained GL
equations).

\begin{figure}
\includegraphics[width=\linewidth]{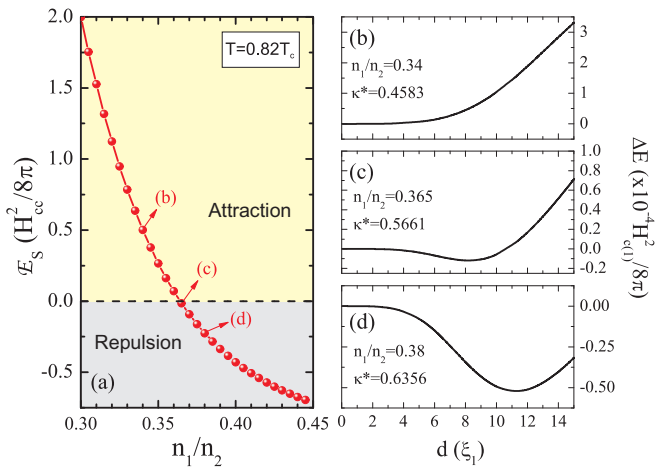}
\caption{(color online) The normal domain - superconductor surface
energy $E_S$ as a function of the ratio of the density of states in
the two bands (a) and the corresponding vortex-vortex interaction
energies (b-d) for indicated parameters. The short-range interaction
force changes sign when the surface energy changes
sign.\label{fig1}}
\end{figure}
As a first example, Fig.~\ref{fig1} shows the surface energy $E_S$
and the numerically obtained vortex-vortex interaction potentials
for a set of parameters corresponding (arguably) to MgB$_2$:
$\kappa_1=3.71$ and $\xi_1/\xi_2 = v_1/v_2$ = 0.255 are taken from
Ref. \onlinecite{Mosh}, the coupling matrix is obtained from Ref.
\onlinecite{gol}, the temperature is fixed at $T = 0.82 T_c$, while
we vary the density of states in the two bands. We note that in all
considered cases $\kappa^*<1/\sqrt{2}$ and the long-range
interaction is always attractive, whereas short-range interaction
changes to repulsive {\it exactly when the surface energy $E_S$
changes sign} with increasing $n_1/n_2$. \textbf{To conclude, the
long-range vortex-vortex interaction is determined by $\kappa^*$
with respect to $1/\sqrt{2}$, while the short-range behavior is
determined by the sign of the surface energy $E_S$}. This also
proves insufficient the initial premise in Ref. \onlinecite{Mosh}
that if the system has $\lambda/\xi_1 > 1/\sqrt{2}$ and
$\lambda/\xi_2 < 1/\sqrt{2}$, the vortex interaction should be
long-range attractive and short-range repulsive. The actual behavior
is far more complex, and can be {\it exactly} determined as
explained above.
\begin{figure}[b]
\includegraphics[width=\linewidth]{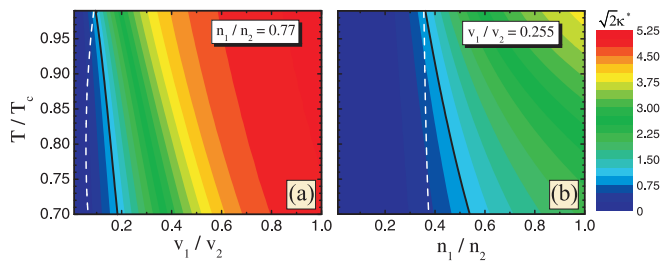}
\caption{(color online) The long-range interaction phase diagram for
a MgB$_2$ crystal, at different temperatures and for varied values
of the ratio between (a) the Fermi velocities, and (b) the partial
density of states, of the two bands. In each panel, the black line
separates the regions of long-range attraction and long-range
repulsion (left to right). The white lines indicate where $E_S$
changes sign and the short-range interaction changes from attractive
to repulsive (left to right). \label{fig2}}
\end{figure}

Recent calculations have shown that as $T \rightarrow T_c$, only
type-I or type-II vortex behavior can be observed.\cite{Geyer}
Indeed, by analyzing $\kappa^*$ and the sign of $E_S$ at $T
\rightarrow T_c$ as explained above, we always found the same type
of interaction in either long- or short-range limit. However, for
$T$ immediately below $T_c$ the sign change of $E_S$ and the
transition from $\kappa^*
> 1/\sqrt{2}$ to $\kappa^* < 1/\sqrt{2}$ occur for different sets of
parameters, opening up the parameter space for observation of the
non-monotonic vortex interaction. This is shown in Fig. \ref{fig2},
where we plot $\kappa^*$ as a function of temperature and also the
ratios between the Fermi velocities (Fig. \ref{fig2}a) and the
partial density of states of each condensate (Fig. \ref{fig2}b). The
black line in Fig. \ref{fig2} denotes $\kappa^* = 1/\sqrt{2}$ and
the white line indicates where $E_S = 0$. At $T = T_c$ these lines
coincide, in agreement with Ref. \onlinecite{Geyer}, but as $T$
decreases, the lines separate, bordering the region where the system
exhibits short-range repulsion ($E_S < 0$) and long-range attraction
($\kappa^* < 1/\sqrt{2}$), i.e. non-monotonic vortex interaction.
This finding further creates a new possibility of {\it tuning the
magnetic interactions} in two-band superconductors {\it by changing
temperature}. For example, for the parameters of MgB$_2$ given in
Ref. \onlinecite{Mosh} (Fig. \ref{fig2}a for $v_1/v_2=0.255$), we
find that non-monotonic vortex interactions occur only for $T
\lesssim 0.49~T_c$,\footnote{This temperature is far out of the
Ginzburg-Landau domain, but is taken as an example. Our prediction
can also be made at a higher temperature, but for microscopic
parameters that do not correspond to any known material at the
moment.} whereas pure type-II behavior is expected at higher
temperatures. The experiment in Ref. \onlinecite{Mosh} was done at
$T \approx 0.1~T_c$, and could thus be repeated at higher
temperatures to verify our prediction.

\begin{figure}[t]
\includegraphics[width=\linewidth]{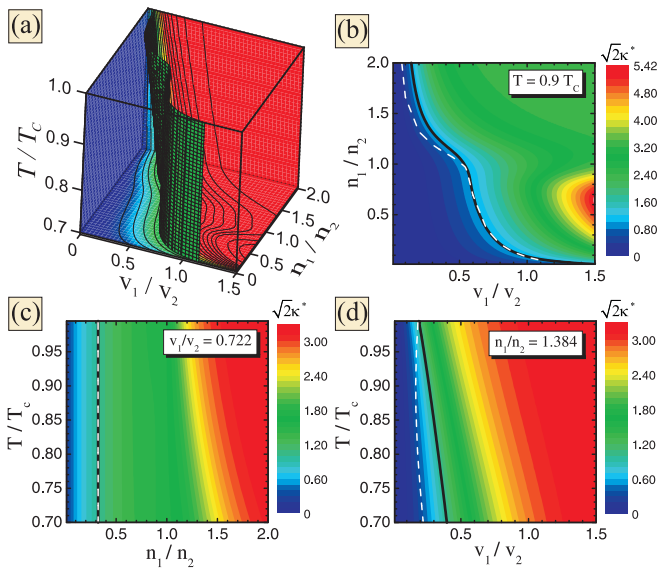}
\caption{(color online) (a) The long-range vortex interaction
$(v_1/v_2,n_1/n_2,T)$ phase diagram for LiFeAs, for other parameters
taken from Ref. \onlinecite{pnic2}. The shown isosurface corresponds to
$\kappa^*=1/\sqrt{2}$ and the change of the long-range vortex-vortex
interaction. (b-d) 2D cuts of (a) in the $T=0.9~T_c$,
$v_1/v_2=0.722$, and $n_1/n_2=1.384$ planes, respectively. Black
(white) lines correspond to $\kappa^*=1/\sqrt{2}$ ($E_S = 0$).
\label{fig3}}
\end{figure}

In Fig.~\ref{fig3}(a) a similar phase diagram is constructed for
recently discovered, and for many reasons exciting, pnictides. In
particular, we show the results for LiFeAs, using the parameters
given in Ref. \onlinecite{pnic2}, except for the fact that
$\lambda_{12}$ in the $\Lambda$ matrix must be taken negative due to
the $s_{\pm}$ pairing. For this material, we extract $\kappa_1 =
2.4$, $n_1/n_2 = 1.384$ and $v_1/v_2 = 0.722$. Interestingly enough,
as $\kappa_2 = \kappa_1\sqrt{n_1v_1^2/n_2v_2^2}$, we note that both
nominal GL parameters of the bands are larger than $1/\sqrt{2}$ if
$\sqrt{n_1v_1^2/n_2v_2^2}\gtrsim 0.295$. Therefore, it can be once
more verified that in a large portion of the parameter space where
both bands are convincingly type-II, the coupled system exhibits
type-I behavior. In Fig.~\ref{fig3}(b), we show that for $T =
0.9~T_c$, the $E_S = 0$ (white) and $\kappa^* = 1/\sqrt{2}$ (black)
curves coincide for small $n_1/n_2$ and large $v_1/v_2$. This
behavior persists even at lower temperatures, as shown in
Fig.~\ref{fig3}(c). However, in the opposite case (large $n_1/n_2$
and small $v_1/v_2$), the curves separate, forming a region of
non-monotonic vortex interaction in the phase diagram which grows
larger as temperature decreases (see Fig.~\ref{fig3}(d)). This broad
temperature range for the observation of partial vortex attraction
is important experimentally, to discriminate the non-monotonic
vortex interactions from irregular vortex lattices formed due to
intrinsic defects in the material \cite{pinn} (with latter being
dominant only at temperatures where the vortex core and the defects
are similar in size, unless defects are of magnetic nature).

\section{Conclusions}
In conclusion, we have demonstrated the semi-analytic method to
relatively easily determine the nature of vortex-vortex interaction
in two-band superconductors. This is of significant theoretical and
experimental importance, as Figs. \ref{fig2}, \ref{fig3} sketch just
two examples of many possibilities attainable by two-band
hybridization. Note that a plethora of transitions, even reentrant
behaviors, can be found as a function of the microscopic parameters,
which can be tuned experimentally (to some extent) by e.g. carrier
injection.\cite{mueller} Finally, with appropriate modifications of
the initial energy functional our approach can also provide insight
in similar situations encountered in nanoscale superconducting
films, tailor-made two-component superconducting hybrids, and dirty
two-band compounds.

\begin{acknowledgments}
Discussions with  A. Moreira, A. Shanenko, R. Prozorov and A. Golubov are
gratefully acknowledged. This work was supported by the Flemish
Science Foundation (FWO-Vl), the Belgian Science Policy (IAP), the
bilateral project FWO-CNPq, CAPES and PRONEX/CNPq/FUNCAP.
\end{acknowledgments}

\end{document}